\newcommand{\slaninafiginline}[1]{}
\newcommand{\slaninafigafter}[1]{#1}
\def\slaninafigsize{ voffset=-420 hoffset=-60 
        hscale=80 vscale=80 }
\def\slaninafigspace{50mm} 
\def\slaninafigsizeh{ voffset=-350 hoffset=-60 
        hscale=80 vscale=80 }
\def\slaninafigspaceh{75mm} 
\documentstyle[prl,aps]{revtex}
\begin{document}
\draft

\twocolumn[\hsize\textwidth\columnwidth\hsize\csname@twocolumnfalse\endcsname

\title{Toy model for the mean-field theory of hard-sphere liquids}

\author{Giorgio Parisi}
\address{Dipartimento di Fisica, Sezione INFN, and Unit\`a INFM,
I Universit\`a di Roma ``La Sapienza'',\\ 
Piazzale Aldo Moro 2, I-00185 Roma, Italy}

\author{Franti\v{s}ek Slanina 
        \cite{f-adr}
}
\address{        Institute of Physics,
	Academy of Sciences of the Czech Republic\\
	Na~Slovance~2, CZ-18221~Praha,
	Czech Republic%
}
\maketitle
\begin{abstract}
We investigate a toy model of liquid, based on simplified HNC
equations in very large spatial dimension $D$. The model does not exhibit
a phase transition, but several regimes of the behavior when
$D\to\infty$ can be observed in
different intervals of the density.  

 \end{abstract}


\twocolumn]

\section{Introduction}
The theory of classical liquids \cite{croxton_74,ba_hen_76} received
recently an important 
stimulus from the the theory of structural glasses 
\cite{me_pa_96,me_pa_99,mez_par_99,col_mez_par_ver_99,col_par_ver_99,mezard_99,do_he_99,co_pa_ve_00,das_va_00}. 
In the pioneering series of papers by Kirkpatrick and Thirumalai
\cite{ki_thi_87,ki_thi_87a} the possible connection of
structural-glass transition with the spin-glass transition in
$p$-spin models was put forward. The analogy was then developed
e. g. in the problem of minimally correlated sequences, which was
shown to possess a glassy behavior without quenched randomness
\cite{ma_pa_ri_94a,ma_pa_ri_94b}. 

However, it would be desirable to put this analogy further.
One of the difficulties occurring in structural glasses, when compared
to spin glasses, comes from the absence of any kind of
analytically solvable mean-field version of the model. In the case of
spin glass, the role is played by the fully-connected Ising model, which
is solvable very easily. The disordered version is the well-known
Sherrington-Kikpatrick model \cite{me_pa_vi_87}, whose understanding is
now very close to  
be complete \cite{young_97}. 

On the contrary, no analytical solution of a ``mean-field'' liquid is
known, as far as we know. It can even sound not very reasonable to
speak about a mean-field liquid, because the relevant high-density
phase is characterized by strong short-range correlations, which can
hardly be replaced by an effective medium. So, the meaning of the mean
field should be better specified. In our investigation, we will understand by
mean-field the situation which occurs in very high dimension,
$D\to\infty$. The purpose of the present work is to
introduce a simple model of a liquid, which is analytically
solvable in the limit of infinite dimension, at least in a certain
well-defined range of densities.

\section{simplified HNC equations}

We consider a liquid composed of hard spheres with the diameter
1. There is only one independent state variable, which is the spatial
density of particles $\rho$.

The configuration of the liquid is described by the radial pair distribution
function $g(r)=h(r)+1$. In the hypernetted chain (HNC) approximation
\cite{meeron_60} 
we have a closed set of equations for the correlation function $h(r)$
\begin{eqnarray}
&&h(r)+1=\exp(W(r)-\beta U(r))\nonumber\\
&&\label{eq:hnc}\\
&&\hat{W}(p)={\rho\hat{h}^2(p)\over 1+\rho\hat{h}(p)}\nonumber\;\;\;\; .
\end{eqnarray}
The potential is $U(r)=0$ for $r>1$ and $U(r)=\infty$ for $r<1$. 
These equations can be interpreted as conditions for minimization of
the free energy functional \cite{me_pa_96}
\begin{eqnarray}
&&{\cal F}\left[h\right]=\nonumber\\
&&\rho^2\int{\rm d}r\; r^{D-1}((h(r)+1)(\ln(h(r)+1)-1+U(r))+1)\\
&&+{1\over (2\pi)^D}\int{\rm d}p\; p^{D-1}L_3(\rho\hat{h}(p))\nonumber
\end{eqnarray}
with
$L_3(x)=-\ln(1+x)+x-\frac{x^2}{2}$.
The function $L_3(x)$ has the following behavior: $L_3(x)\to\infty$
for $x\to -1$ and $L_3(x)\simeq -x^3/3$ for $x\ll 1$.

Our main approximation will consist in replacement the 
function $L_3(x)$ by $L_\infty(x)$, where $L_\infty=\infty$ for $x<-1$
and $L_\infty=0$  otherwise.  The motivation for this approximation is that we suppose that the main efffect of $L_3(x)$ is to forbid the region, where $-\rho\hat{h}(p)>1$. Then, minimization of the free
energy functional amounts to satisfying the conditions
\begin{eqnarray}
\rho \hat{h}(p)&\ge& -1\nonumber\\
h(r)&\ge& -1
\label{eq:condforh}\\
h(r)&=&-1\;{\rm for}\; r<1\nonumber
\end{eqnarray}
which are in fact the minimum physical requirements for any
correlation function $h(r)$. In this sense we are building a
``minimum'' model of a liquid. 
In addition to the constraints (\ref{eq:condforh}) we require  
that the function $h(r)$ depends continuously on the density.    
The absence of a solution which would be continuous in density would
be a signal of a phase transition. This is, however, not found in the
present calculations.

\slaninafigafter{
\begin{figure}[hb]
  \centering
  \vspace*{\slaninafigspaceh}
  \includegraphics{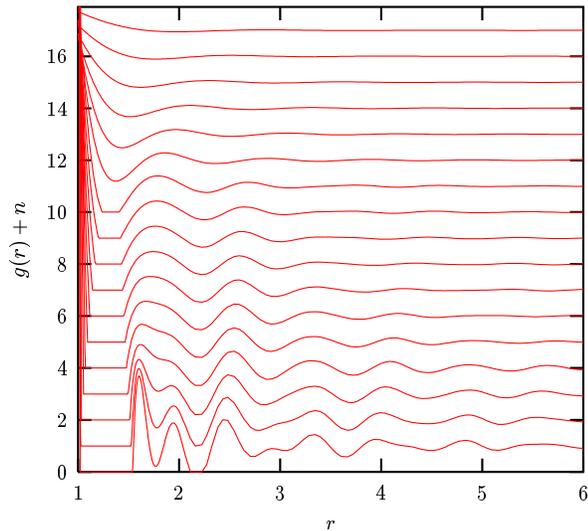}
  \caption{Pair distribution function $g(r)=h(r)+1$ of the  3-dimensional
           model liquid,
           for densities (from top to bottom)
           $\rho=0.6$, $0.7$, $0.8$, $0.9$, $1.00$, $1.1$, $1.2$,
  $1.23$, $1.26$, $1.29$, $1.32$, $1.35$, $1.38$, $1.41$, $1.44$,
  $1.47$, $1.5$. The $n$-th curve    
           from the bottom is shifted by $n$ upwards.}
  \label{fig:gofr-all}
\end{figure}
}

In 3D we can compute the function $h(r)$ numerically by increasing
slowly the density $\rho$ and adjusting iteratively the function
$h(r)$ so as the conditions (\ref{eq:condforh}) are satisfied. The
resulting pair distribution function $g(r)=h(r)+1$ is shown in the
Fig. \ref{fig:gofr-all}. The Fourier transform $\hat{h}(p)$ for
$\rho=1.2$ is shown in Fig. \ref{fig:strucfac}.

\slaninafigafter{
\begin{figure}[hb]
  \centering
  \vspace*{\slaninafigspace}
  \includegraphics{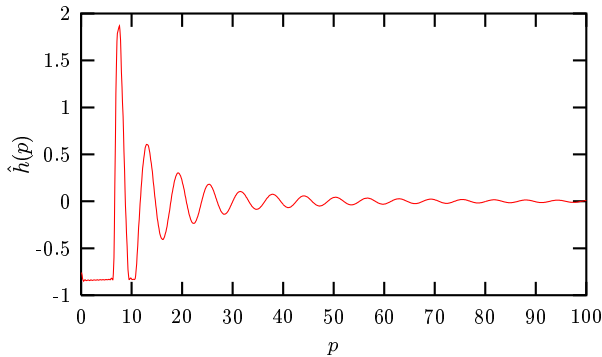}
  \caption{Fourier transform of the correlation function for
  3-dimensional liquid, at the density $\rho=1.2$.
           }
  \label{fig:strucfac}
\end{figure}
}

We can see that for densities up to about
$\rho=1$  the pair distribution function agrees qualitatively with the
well-known results of HNC approximation or numerical simulations (see
\cite{croxton_74}). However, at about $\rho=1.2$ the behavior
changes. A gap opens between the principal peak at $r=1$ and the
secondary peak at $r\simeq 2$. the gap broadens with increased density
and at about $\rho \simeq 1.5$ a second gap occurs around 
$r\simeq 2.2$. We observed, that further compression leads to the
occurrence of a third  gap separating the peaks at $r\simeq 1.6$ and
$r\simeq 2$. We expect that continued increase of the density results
in increased number of gaps. 

The presence of the gaps is an artifact of the approximation. In
reality the values of $g(r)$ will not be strictly zero, but small. 

From the value of the radial distribution function at $r=1$ the
pressure can be computed \cite{croxton_74} and the resulting equation
of state is shown in Fig. \ref{fig:pressure}, together with the
results obtained by solving the HNC equations (\ref{eq:hnc}) and the
formula computed 
in the Percus-Yevick (PY) approximation \cite{croxton_74}.
We can see that our model behaves qualitatively in the same way as the
other approximations, even though quantitative agreement is poor. On
the other hand, the equation of state of our model does not differ
from either HNC or PY approximation more than these two approximations
differ one from the other.

\slaninafigafter{
\begin{figure}[hb]
  \centering
  \vspace*{\slaninafigspace}
  \includegraphics{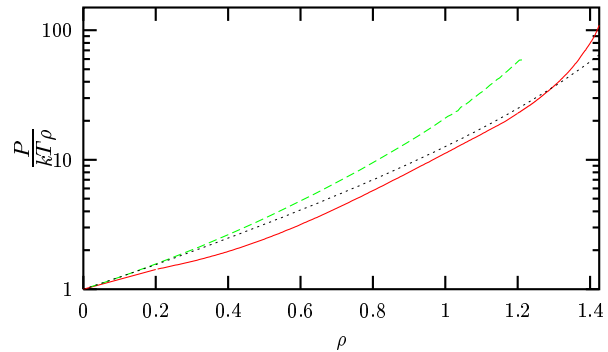}
  \caption{
  Equation of state for 3-dimensional liquid. Our model: full 
  line (red).  HNC approximation: dashed line (green). Percus-Yevick
  approximation: dotted line (black).}
  \label{fig:pressure}
\end{figure}
}

We can see from these results, that the present approach in 3D gives at
least qualitatively sensible results. However, our aim is not to
provide a new approximation for real three-dimensional liquids, but a
model which describes reasonably well the qualitative features of a liquid 
and is soluble in the limit of infinite spatial dimension. This will
be done in the next section.

\section{Solution of the model in high dimension}

In this section we will investigate a $D$-dimensional version of the
model, with $D=2N+3$ and $N\to\infty$.
The main quantity of interest is again the correlation function $h(r)=g(r)-1$.
We will suppose that it has the form
$h(r)=h_0(r)+\bar{h}(r)$
where
$h_0(r)=-\theta(1-r)$
and $\bar{h}(r)=0$ for $r<1$. 

The pressure is directly related to the value of that $r=1$, more precisely to 
$\lim_{r\to 1^+}h(r)=\bar{h}(1^+)$ (see \cite{croxton_74})
\begin{equation}
\frac{1}{kT}P=\rho+\frac{1}{2}V_D\rho^2(1+\bar{h}(1^+))\;\; .
\end{equation}
In the following, we will use rescaled quantities. We will use 
$\bar{\rho}=\rho V_D$ for the density, while for the pressure we use
$\bar{P}={V_DP\over kT}$.

In order to check the conditions (\ref{eq:condforh}) we should know the
properties of the Fourier transform in high dimension. The relevant
formulae used in this section are given in the Appendix A. 

In the zeroth approximation
$\hat{h}(\hat{p})=-V_D\;\Psi(\hat{p})$, which is correct as long as
$\rho < 1/V_D$. 
So, for $\bar{\rho}<1$ we have $\bar{h}(r)=0$ and the pressure
is given by first virial correction, 
\begin{equation}
\bar{P}=\bar{\rho}+\frac{1}{2}\bar{\rho}^2\;\; .
\label{eq:virial}
\end{equation}

For $\bar{\rho}>1$ 
we rewrite $h$ in the form
\begin{equation}
h(r)=h^*_0(r)+\bar{h}^*(r)
\end{equation}
where still $\bar{h}^*=\bar{h}(r)$ for $r>1$, but $\bar{h}^*(r)$ is
continuous for all $r$. The Fourier transform of $h^*_0$ is easy
to compute, if we know the value $h^*(1^-)=-1-A$, where
$A=\bar{h}(1^+)=\hat{h}_1(1)$.  Then
\begin{equation}
\widehat{h^*_0}(\hat{p})=-(1+A)\,V_D\,\Psi_0(\hat{p})\;\; .
\end{equation}
In order to ensure $\hat{h}(\hat{p})\ge 1/\rho$ everywhere, we set
\begin{equation}
\widehat{\bar{h}^*}(\hat{p})=
\theta_1 \left((A+1)V_D\,\Psi_0(\hat{p})-\frac{1}{\rho}\right)\;\; .
\end{equation}
(We denote $\theta_1(x)=x$ for $x>0$, $\theta_1(x)=0$ for $x\le 0$.)

The relevant quantity is 
\begin{equation}
\rho_1={\ln \bar{\rho}\over N}\;\; .
\end{equation}
Indeed, one particle occupies space $2^{-D}V_D$, so an absolute upper
bound for the density  is $\rho_1<\ln 4$.

Let us suppose that $\hat{p}_c < 1$. This condition restrict the range
of densities investigated to a certain interval, which will be found
in what follows.

For $\hat{p}<1$ the function
$\Psi_0(\hat{p})=\exp(-N\phi_0(\hat{p}))$ is monotonously decreasing 
and we have the following
equation for $\hat{p}_c$. 
\begin{equation}
\phi_0(\hat{p}_c)=\rho_1-{\ln(A+1)\over N}
\label{eq:forpc}
\end{equation}

When computing the inverse Fourier transform of
$\hat{\bar{h}^*}(\hat{p})$ we need only the behavior around the point
$\hat{p}_c$, which is
\begin{equation}
\hat{\bar{h}^*}(\hat{p})\simeq
-V_D\,\Psi_0^\prime(\hat{p}_c)\,\hat{p}_c\,(1-{\hat{p}\over\hat{p}_c})
\end{equation}
and gives, using (\ref{eq:invfourlin})
\begin{equation}
\hat{h}_1(r)=\left({N\over 2\pi}\right)
{V_D^2\,\hat{p}_c^{D+1}(-\Psi_0^\prime(\hat{p}_c))(A+1)
\over D+1}\,\Psi_0(\hat{p}_c r)
\end{equation}

We obtain immediately the equation for $A$
\begin{equation}
A=\left({N\over 2\pi}\right)^D
{N\over D+1}
{V_D^2\,\hat{p}_c^{D+1}\phi_0^\prime(\hat{p}_c))
(A+1)\Psi_0^2(\hat{p}_c)}\; .
\label{eq:forA}
\end{equation}

In the interval 
$\hat{p}_c<1$
we have
\begin{equation}
\phi_0^\prime(\hat{p}_c)={\hat{p}_c\over 1+\sqrt{1-\hat{p}^2_c}}<\hat{p}_c\;\; .
\end{equation}
Hence, if we suppose that a solution such that $A<1$ exists, we have
\begin{equation}
A\le \left({N\over 2\pi}\right)^D 
V_D^2 \hat{p}_c^{D+2}\Psi_0^2(\hat{p}_c)
\simeq \exp(-2NK(\hat{p}_c))
\end{equation}
where
\begin{equation}
K(\hat{p})=\ln(1+\sqrt{1-\hat{p}^2})-\sqrt{1-\hat{p}^2}-\ln\hat{p}\;\; .
\end{equation}
We have $K(1)=0$ and $K(\hat{p})=-\sqrt{1-\hat{p}^2}\;/\hat{p}<0$, so
$K(\hat{p})>0$ for $\hat{p}<1$. For fixed $\hat{p}_c<1$ and
$N\to\infty$ we have $A\ll 1$ and therefore we can neglect $A$ in the
equations (\ref{eq:forpc}) and (\ref{eq:forA}).

So, we can conclude that in the range of densities
$\rho_1<\rho_{1c}=\phi_0(1)=1-\ln(2)=0.3068...$ the following equation for
$\hat{p}_c$ holds
\begin{equation}
\rho_1=\ln(1+\sqrt{1-\hat{p}_c^2})-\sqrt{1-\hat{p}_c^2}+1-\ln 2\; .
\label{eq:forpcunder1}
\end{equation}

The solution of the latter equation can be easily obtained in the
form of power series. We show here only first several terms. The
expansion up to order 16 is given in the Appendix B and the graph is
shown in Fig. \ref{fig:pcofrho}.
\begin{equation}
\hat{p}_c^2=
4\,\rho_1-2\,{\rho_1}^{2}-\frac{2}{3}\,{\rho_1}^{3}-\frac{5}{6}\,{\rho_1}^{4}-O({\rho_1}^{5})\; .
\end{equation}
\slaninafigafter{
\begin{figure}[hb]
  \centering
  \vspace*{\slaninafigspace}
  \includegraphics{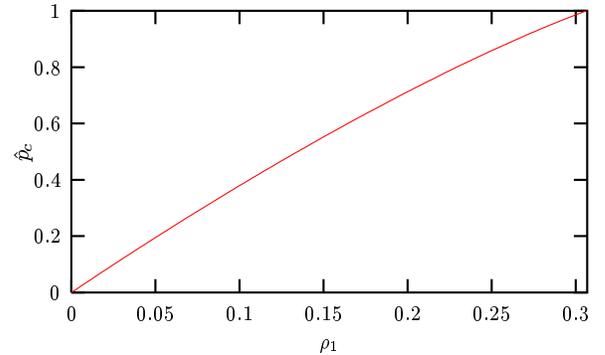}
  \caption{Dependence of the momentum $\hat{p}_c$ on the rescaled
  density in the regime $\hat{p}_c<1$.
           }
  \label{fig:pcofrho}
\end{figure}
}

From the solution Eq. (\ref{eq:forpcunder1}) we can compute the pressure. 
We write $\bar{P}=\bar{\rho}+\frac{1}{2}\bar{\rho}^2+P_c$
and re-scale the correction as $P_c=\exp(NP_1)$. We obtain
\begin{equation}
P_1=2-2\ln 2 +\ln \hat{p}_c^2
\label{eq:pressurecorr}
\end{equation}
We can see that the density  $\rho_{1t}=\phi_0\left({2\over {\rm
e}}\right)=0.14676...$ separates two regimes. For
$\rho_1<\rho_{1t}$ the correction $P_c$  to the lowest virial becomes
negligible for large $N$, while for $\rho_1>\rho_{1t}$ the
correction diverges for  $N\to\infty$.
The graph of the function $P_1(\rho_1)$ is shown in Fig. \ref{fig:p1ofrho}.

\slaninafigafter{
\begin{figure}[hb]
  \centering
  \vspace*{\slaninafigspace}
  \includegraphics{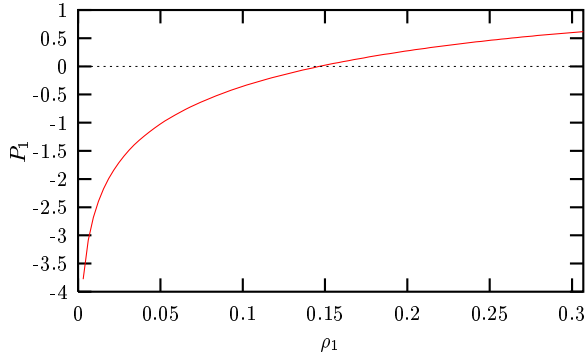}
  \caption{Equation of state for our model in the limit of infinite
  dimension, in the range of densities where $\hat{p}_c<1$.}
  \label{fig:p1ofrho}
\end{figure}
}

For the correlation function  in the interval
$r\hat{p}_c<1$ 
we have $h(r)=\exp(N h_1(r))$, where
\begin{eqnarray}
&&h_1(r)=\nonumber\\
&&1-2\ln 2 + \ln \hat{p}_c^2 -\rho_1+\\
&&+\sqrt{1-\hat{p}_c^2 r^2}
-\ln(1+\sqrt{1-\hat{p}_c^2 r^2})
\;\;.
\nonumber
\end{eqnarray}
and for $r\hat{p}_c>1$ we 
can use the following scaling
$\hat{h}(r)=\exp(Nh_1(r))\,\cos(Nh_2(r))$
where, using the expressions (\ref{eq:phi1}) and (\ref{eq:phi2}), we obtain
\begin{equation}
h_1(r)=1-2\ln 2 + \ln \hat{p}_c -\rho_1-\ln r
\end{equation}
\begin{equation}
h_2(r)=\sqrt{\hat{p}_c^2 r^2 -1}-\arctan\sqrt{\hat{p}_c^2 r^2-1}\;\; .
\end{equation}
We can also see that $|h_1(r)\ll 1$ for $\rho_1<\rho_{1c}$, so that
$g(r)>0$ for $r>1$ and the gaps in $g(r)$, discussed in the last section
do not occur. However, when $\rho_1$ approaches
$\rho_{1c}$ the absolute value of $h_1(r)$ can be of order $O(1)$ and
a gap can appear at the density $\rho_{1c}$. The
detailed investigation of this process and the behavior of the model
for $\rho_1>\rho_{1c}$ remains an open question.

\section{Conclusions}
We investigated a simple model for a hard-sphere liquid. By numerical
solution in 3 dimensions, we found a qualitatively realistic
behavior. The results for the equation of state
are compatible with the hypernetted chain and Percus-Yevick
approximations. 

We solved the model analytically 
in the limit of large spatial dimension. We found that  two scales of
density and pressure appear, which corresponds to two regimes of density.
For $\bar{\rho}<1$ the equation of state is given by first virial correction
(\ref{eq:virial}), while for $\bar{\rho}>1$ the quantity relevant to
the further virial corrections is $\rho_1=\ln \bar{\rho}/N$ and the
pressure correction itself scale as $P_1=\ln P_c/N$ (\ref{eq:pressurecorr}).
We have found the solution in the interval
$0<0\rho_1<\rho_{1c}=0.3068...$.
Two regimes are present within this interval. For $\rho_1<\rho_{1t}=0.14676...$
the correction $P_c$ vanishes for large $N$, while for
$\rho_1<\rho_{1t}$ it diverges for large $N$. 

It should be expected, that the
presence of presence of gaps in $g(r)$ will lead
to qualitatively different behavior for densities higher than $\rho_{1c}$.

\acknowledgments{One of us (F.S.) wishes to thank to the INFN section of Rome
University ``La Sapienza'' for financial support and kind hospitality.
This work was supported by the project No. 202/00/1187 of the Grant Agency
of the Czech Republic.
}

\section*{Appendix A}
Here we derive the formula for the Fourier transform in high dimension.

Fourier transform in $D=2N+3$ dimensions is ($x$ and $k$ are
$D$-dimensional vectors) 
\begin{equation}
\hat{f}_D(k)=\int{\rm d}^D x\; {\rm e}^{{\rm i}xk}f_D(x)
\end{equation}
\begin{equation}
f_D(x)=(2\pi)^{-D}\int{\rm d}^D k\; {\rm e}^{{\rm -i}xk}\hat{f}_D(k)\;\; .
\end{equation}
We suppose that the functions depend only on the radial coordinate,
$f(r)=f_D(x)$ for $r=|x|$ and $\hat{f}(p)=\hat{f}_D(k)$ for $p=|k|$.
After rescaling $\hat{p}=p/N$ we will finally have
\begin{eqnarray}
&&\hat{f}(\hat{p})=\nonumber\\
&&C_N\int_0^\infty{\rm d}r\int_{-1}^1{\rm d}z\;
\left[ r^{2(1+\frac{1}{N})} (1-z^2){\rm e}^{{\rm i}\hat{p}rz}
\right]^N f(r)
\end{eqnarray}
where the constant $C_N$ is fixed by condition that for
$f(r)=\theta(1-r)$ we have $\hat{f}(0)=V_D$ with $V_D$ volume of the
$D$-dimensional sphere,
\begin{equation}
V_D={2\pi^{D/2}\over D\Gamma(\frac{D}{2})}
\simeq \left({{\rm e}\pi\over N}\right)^N\;\; .
\end{equation}
Similarly for the inverse Fourier transform we will have
\begin{eqnarray}
&&f(r)=\nonumber\\
&&\hat{C}_N\int_0^\infty{\rm d}\hat{p}\int_{-1}^1{\rm d}z\;
\left[ \hat{p}^{2(1+\frac{1}{N})} (1-z^2){\rm e}^{{\rm i}\hat{p}rz}
\right]^N \hat{f}(\hat{p})
\end{eqnarray}
with coefficient
\begin{equation}
\hat{C}_N=C_N \left({N\over 2\pi}\right)^D\;\; .
\end{equation}
The calculation of the Fourier transform 
can be performed by the saddle-point
method. The essential result is the Fourier transform of the surface of unit sphere, $f(r)=\delta(r-1)$. We obtain $\hat{f}(\hat{p})\propto\Psi(\hat{p})$
where
\begin{equation}
\Psi(\hat{p})=\Psi_0(\hat{p})=\exp(-N\phi_0(\hat{p}))
\end{equation}
for  $\hat{p}<1$ and 
\begin{equation}
\Psi(\hat{p})=\Psi_1(\hat{p})=\exp(-N\phi_1(\hat{p}))\cos(N\phi_2(\hat{p}))
\end{equation}
for $\hat{p}>1$.

The explicit form of the functions $\phi_0,\phi_1,\phi_2$ is given by
\begin{eqnarray}
&&\phi_0(\hat{p})=
1-\ln 2 + \ln(1+\sqrt{1-\hat{p}^2})-\sqrt{1-\hat{p}^2}\\
&&\phi_1(\hat{p})=1-\ln 2 + \ln\hat{p}\label{eq:phi1}\\
&&\phi_2(\hat{p})=\sqrt{\hat{p}^2-1}-\arctan\sqrt{\hat{p}^2-1}\;\; .
\label{eq:phi2}
\end{eqnarray}
Note that $\Psi(0)=1$.

From here we can deduce the following Fourier transforms 
($\theta(x)=1$ for $x>0$ and $\theta(x)=0$ for
$x<0$).
For $f(r)=A\;\theta(r_0-r)$:
\begin{equation}
\hat{f}(\hat{p})=AV_D\;r_0^D \; \Psi(\hat{p}r_0)\;\; .
\end{equation}
For $f(r)=A\;(1-r/r_0)\;\theta(r_0-r)$:
\begin{equation}
\hat{f}(\hat{p})={AV_D\;r_0^D \over D+1} \; \Psi_0(\hat{p}r_0)\;\; .
\end{equation}

Because the inverse Fourier transform has the same form and differs
only in the factor $\hat{C}_N$ instead of $C_N$, we can also
immediately write
for $\hat{f}(\hat{p})=A\;\theta(\hat{p}_c-\hat{p})$:
\begin{equation}
f(r)=\left({N\over 2\pi}\right)^DAV_D\;\hat{p}_c^D\;
\Psi(\hat{p}_c r)
\end{equation}
and for $\hat{f}(\hat{p})=A\;(1-\hat{p}/\hat{p}_c)\;\theta(\hat{p}_c-\hat{p})$
\begin{equation}
f(r)=\left({N\over 2\pi}\right)^D {AV_D\;\hat{p}_c^D \over D+1}\;
\Psi(\hat{p}_c r)\;\; .
\label{eq:invfourlin}
\end{equation}

\section*{Appendix B}

Using Maple V we get the following expansion for the solution of
equation (\ref{eq:forpcunder1}).  
\begin{eqnarray}
&&\hat{p}_c^2=\nonumber\\
&&4\,\rho_1-2\,{\rho_1}^{2}-2/3\,{\rho_1}^{3}-5/6\,{\rho_1}^{4}
-{\frac {41}{30}}
\,{\rho_1}^{5}-{\frac {469}{180}}\,{\rho_1}^{6}-\nonumber\\
&&\nonumber\\
&&-{\frac {6889}{1260}}\,{
\rho_1}^{7}-{\frac {24721}{2016}}\,{\rho_1}^{8}-{\frac {2620169}{90720}}\,
{\rho_1}^{9}-{\frac {64074901}{907200}}\,{\rho_1}^{10}-
\nonumber\\
&&\nonumber\\
&&-{\frac {1775623081}
{9979200}}\,{\rho_1}^{11}-
-{\frac {1571135527}{3421440}}\,{\rho_1}^{12}-\nonumber\\
&&\\
&&-
\frac {1882140936521}{1556755200}\,{\rho_1}^{13}
-{\frac {70552399533589
}{21794572800}}\,{\rho_1}^{14}-\nonumber\\
&&\nonumber\\
&&-{\frac {2874543652787689}{326918592000}}
\,{\rho_1}^{15}-{\frac {25296960472510609}{1046139494400}}\,{\rho_1}^{16}
\nonumber
\;\; . \nonumber
\end{eqnarray}

\end{document}